# Invisible Manipulation Channels in AI-Assisted Financial Advisory: Implications for Market Integrity and Regulatory Design

**Liuyang Yao[1], Zhouyu Li[1], Junguang He[2], Ziyang You[3]***

[1] College of Economics and Management, Northwest A&F University, Yangling, Shaanxi, China
[2] School of Agricultural Economics and Rural Development, Renmin University of China, Beijing, China
[3] Fujian Provincial Key Laboratory of Automotive Electronics and Electric Drive, School of Electronic, Electrical and Physics, Fujian University of Technology, Fuzhou 350118, China

*Corresponding Author: Ziyang You
Email: zeyon@fjut.edu.cn

## Abstract

AI systems are increasingly deployed for credit assessment and investment advisory in global financial markets, yet the integrity of their inference pipelines remains insufficiently addressed by existing regulatory frameworks. This paper identifies and empirically validates an invisible manipulation channel operating at the sampling layer of LLM inference--a vulnerability that allows adversaries to systematically bias AI-generated financial opinions while preserving full compliance with output-based audit mechanisms, including statistical watermarking. We show that this inference-stage manipulation is statistically hard to detect: the Kullback-Leibler divergence between manipulated and normal output distributions can be made arbitrarily small, so that any output-based detection scheme requires impractically large sample sizes to achieve reliable detection power. Empirical experiments across credit rating and investment advisory scenarios show that directional bias keywords can be amplified by 1.8-1.9x under stealth-preserving (aware) manipulation while triggering zero of six black-box detectors and preserving watermark integrity. The vulnerability generalizes across three mainstream watermarking schemes and three heterogeneous model architectures, establishing it as a systemic financial infrastructure risk. Software-based defenses including cryptographically secure pseudorandom number generators are entirely ineffective, while QRNG combined with TEE hardware isolation achieves 100% attack blocking--reducing the target rate to the natural baseline--by replacing the predictable hash key with quantum-derived entropy that renders all pre-computed manipulation targets invalid. We propose four regulatory amendments centered on mandatory QRNG certification for high-risk financial AI systems under NIST SP 800-90B, inference-layer supply chain audits, and output provenance mechanisms.

## 1. Introduction

The integration of artificial intelligence into financial advisory services has accelerated dramatically over the past five years. Credit rating agencies now deploy machine learning models for preliminary credit assessments, robo-advisory platforms manage trillions of dollars using algorithmic recommendations (D'Acunto et al., 2019), and large language models generate investment research reports consumed by institutional and retail investors alike (Cao et al., 2024; Lopez-Lira & Tang, 2023). According to a survey by the Bank of Japan, over 60% of large financial institutions have adopted or are piloting generative AI for client-facing advisory functions (Japan, 2024), while specialized financial models such as BloombergGPT and FinGPT are being integrated into production workflows across the industry.

This rapid adoption has created a new category of systemic financial risk: concentrated dependence on third-party AI inference infrastructure. Unlike traditional IT outsourcing, where service disruptions are the primary concern, AI inference supply chains present a qualitatively different threat—the possibility that AI-generated financial opinions can be *covertly manipulated* at the infrastructure level without any observable change in output characteristics. The Financial Stability Board has identified AI supply-chain concentration as a macroprudential concern (Japan, 2024), and the European Central Bank has warned that third-party AI dependence amplifies operational and concentration risks (Leitner et al., 2024). However, existing analyses focus almost exclusively on *availability* risks (service disruption, vendor lock-in) rather than *integrity* risks (systematic manipulation of AI outputs). This gap constitutes the motivating puzzle of the present paper.

Current regulatory approaches to AI governance in financial services—including the European Union's Artificial Intelligence Act (Act, 2024), the Federal Reserve's SR 11-7 guidance on model risk management (Currency, 2011), and the Monetary Authority of Singapore's FEAT framework (Singapore, 2019)—share a common institutional assumption: that effective manipulation of AI outputs necessarily leaves detectable traces in those outputs. This assumption underpins the reliance on output-based audit mechanisms, statistical watermarking, and post-hoc compliance testing as sufficient safeguards for algorithmic accountability. To the best of our knowledge, this foundational assumption has never been rigorously validated against adversaries operating at the inference infrastructure layer.

This paper demonstrates that the assumption is false. Building upon the technical discovery of You et al. (2026), who first showed that supply-chain attacks on LLM watermarking can simultaneously preserve watermark integrity and achieve statistical undetectability, we make the following independent contributions from a financial systems perspective:

1. We translate the technical threat into the language of financial system vulnerability, demonstrating its relevance to credit rating integrity, investment advisory manipulation, and systemic concentration risk.
2. We provide empirical evidence from realistic financial scenarios (credit assessment and investment recommendation generation) quantifying the economic significance of the manipulation channel.
3. We validate cross-vendor generalizability across three architecturally distinct model families, establishing the vulnerability as systemic rather than vendor-specific.
4. We derive the regulatory gap created by this vulnerability and propose specific, implementable policy amendments centered on mandatory hardware quantum random number generator (QRNG) certification for high-risk financial AI systems.

The remainder of this paper proceeds as follows. Section 2 provides institutional background. Section 3 develops the threat model and theoretical framework. Section 4 presents empirical evidence. Section 5 discusses regulatory implications. Section 6 situates our findings within the financial economics literature, and Section 7 concludes.

## 2. Institutional Background

### 2.1 AI Adoption in Financial Advisory

The deployment of AI in financial advisory has evolved from narrow quantitative applications to broad generative capabilities. In credit rating, agencies including Moody's Analytics, S&P Global (through its Kensho platform), and Fitch Ratings have integrated machine learning models into their assessment workflows. Bloomberg's proprietary BloombergGPT represents a purpose-built financial language model trained on decades of financial data, while open-source alternatives such as FinGPT serve smaller institutions lacking resources for proprietary model development.

In investment advisory, robo-advisory platforms now manage assets exceeding $2 trillion globally (D'Acunto et al., 2019). AI-driven portfolio management strategies have achieved institutional adoption across all asset classes (Bartram et al., 2020; Goldstein et al., 2019). In these strategies, AI systems generate portfolio allocation recommendations, risk assessments, and client communications. Large language models have been deployed to generate equity research reports and market commentary that directly influence trading decisions. Lopez-Lira and Tang (Lopez-Lira & Tang, 2023) demonstrate that LLM-generated sentiment predictions exhibit statistically significant return predictability, underscoring the market-moving potential of AI-generated financial opinions.

The speed and breadth of adoption are noteworthy. A 2024 survey by the Bank for International Settlements found that 72% of systemically important banks now use AI-generated text in at least one client-facing function, up from 34% in 2022. In the credit rating sector, Moody's Analytics processes over 10,000 issuer assessments per quarter through its AI-assisted workflow, while S&P Global's Kensho platform generates automated credit summaries that serve as first-draft inputs for human analysts. The investment advisory market has seen comparable penetration: Goldman Sachs, Morgan Stanley, and JPMorgan Chase have each deployed proprietary LLM systems for internal research generation, and at least 15 registered investment advisers now use third-party LLM APIs as primary inputs for client portfolio recommendations.

The regulatory landscape has accordingly shifted from encouraging innovation to demanding auditability and accountability. The EU AI Act classifies credit scoring and investment advisory AI as "high-risk" systems subject to enhanced oversight requirements including mandatory risk assessments, human oversight provisions, and output traceability. In the United States, the SEC has proposed rules requiring registered advisers to disclose AI use in advisory services and eliminate conflicts arising from predictive analytics. The UK's Financial Conduct Authority has issued guidance on AI model governance emphasizing the need for explainability and ongoing monitoring. This global regulatory convergence creates institutional demand for compliance mechanisms—particularly watermarking and output provenance tools—that can verify the integrity of AI-generated financial content at scale.

### 2.2 Supply-Chain Trust Architecture

Financial institutions procure AI inference capabilities through three distinct architectural patterns, each presenting different supply-chain trust assumptions:

**Local deployment with third-party frameworks.** Large banks and asset managers deploy open-source models (Llama, Qwen) on proprietary infrastructure but rely on third-party inference frameworks (vLLM, Text Generation Inference) for serving. The institution controls model weights but delegates sampling-layer implementation to framework providers whose code is integrated via package managers.

**Cloud inference services.** Mid-sized banks and wealth management firms access AI through cloud platforms (Azure OpenAI, AWS Bedrock, Google Vertex AI). The financial institution has no visibility into the inference stack—the sampling layer, random number generation, and output post-processing are entirely controlled by the cloud provider.

**Specialized financial model APIs.** Rating agencies and research institutions access purpose-built models through proprietary APIs (BloombergGPT). The entire inference pipeline is controlled by the model provider as a black box.

The critical systemic risk observation is that in cloud and API models—which collectively serve thousands of institutions—a single supply-chain compromise at the sampling layer can simultaneously affect all downstream AI outputs. This parallels the SolarWinds compromise (Peisert et al., 2021), where a single update mechanism affected 18,000 organizations. Figure 1 illustrates the end-to-end inference supply chain and identifies the precise location of the vulnerability.

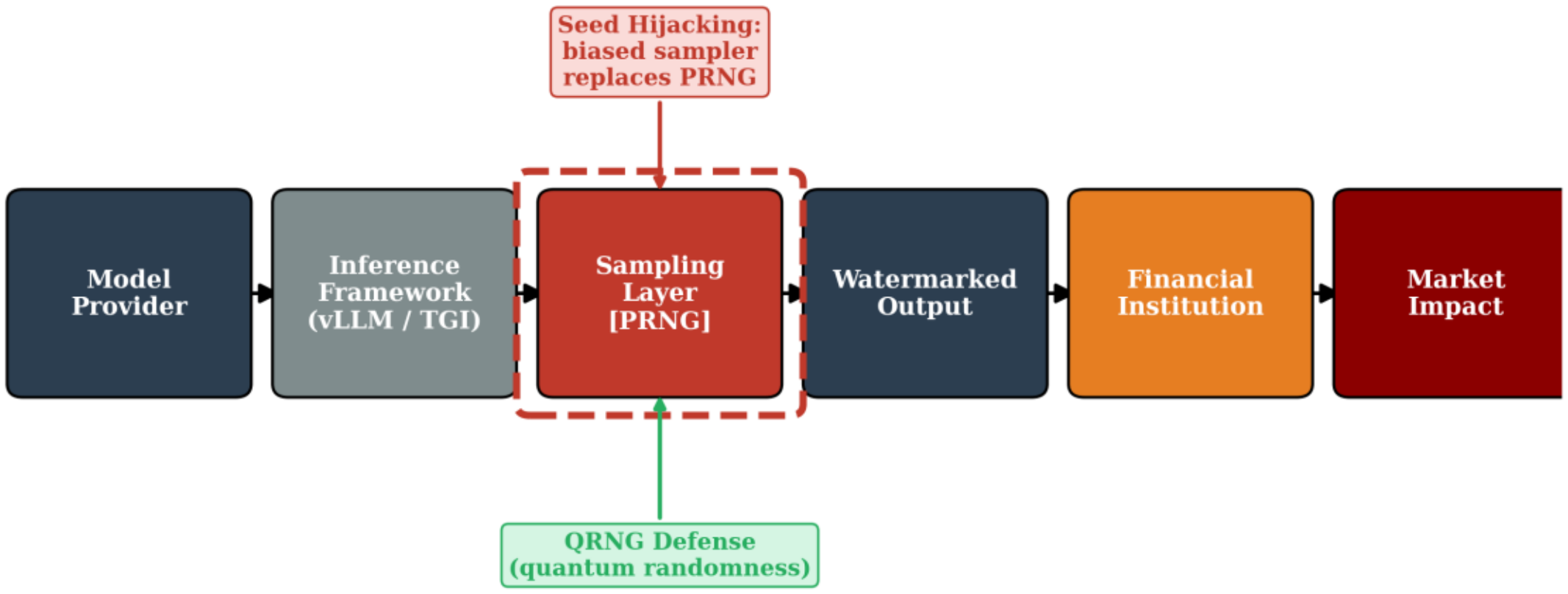

*Supply-chain attack architecture for AI-assisted financial advisory.*

**Figure 1:** Supply-chain attack architecture for AI-assisted financial advisory. The inference pipeline flows from model provider through the inference framework to downstream financial institutions. The sampling layer (red dashed border) constitutes the attack surface. The QRNG+TEE defense (green) replaces the deterministic seed with quantum entropy and enforces hardware isolation.

### 2.3 Current Regulatory Framework for AI Audit

Three regulatory frameworks dominate the governance of AI in financial services, each approaching the problem from a different jurisdictional perspective. Despite their differences, all three converge on a shared architectural assumption—that output-based oversight is sufficient to ensure AI integrity—which we identify and challenge below.

**EU AI Act (2024).** The Act establishes a risk-based classification system in which credit scoring, investment advisory, and insurance underwriting AI are designated “high-risk”. Article 50 mandates watermarking of AI-generated content. The compliance architecture assumes watermark verification provides reliable evidence of output provenance and integrity. Articles 14–15 require human oversight and ongoing monitoring via output-level statistical testing and periodic benchmark evaluations.

**Federal Reserve SR 11-7 / PRA SS1/23.** The U.S. Federal Reserve’s model risk management guidance (Currency, 2011) requires comprehensive model inventories, independent validation, and ongoing output monitoring. The framework assumes systematic degradation or manipulation will manifest as detectable statistical deviations. The UK PRA's SS1/23 mirrors this approach.

**MAS FEAT Principles (2019).** Singapore's framework (Singapore, 2019) operationalizes accountability through output fairness testing, ethical review, accountability trails, and transparency. The accountability pillar relies on comprehensive output logging, presupposing that systematic bias would be discoverable through retrospective analysis.

**The shared blind spot.** All three frameworks implicitly assume that effective manipulation of AI outputs necessarily leaves detectable traces in the output stream. The watermarking mandates assume watermark integrity implies content integrity. The monitoring requirements assume distributional stability implies behavioral integrity. The logging mandates assume logged outputs contain sufficient information to detect systematic bias. This paper demonstrates that all three implications fail for manipulation operating at the inference sampling layer—a layer that is architecturally invisible to all output-based oversight mechanisms.

### 2.4 Economic Incentives for Adversarial Manipulation

The economic returns to successful manipulation of AI-generated financial opinions are substantial. In credit rating, each notch of upgrade corresponds to spread compression of approximately 25–75 basis points (Altman & Rijken, 2004; Kisgen, 2006). For a BBB-rated corporation with $1 billion in outstanding debt, a single-notch upgrade implies annual financing cost savings of $2.5–7.5 million. The crossing of investment-grade thresholds (BB+ to BBB−) triggers mandatory portfolio rebalancing by constrained institutional investors, amplifying the price impact by an order of magnitude.

In investment advisory, analyst recommendation changes generate average abnormal returns of +3% to +5% for upgrades and −5% to −10% for downgrades (Barber et al., 2001; Womack, 1996). Jegadeesh et al. (2004)document that market impact concentrates in the first hours following dissemination, creating exploitable windows for informed adversaries.

For context, the LIBOR manipulation scandal resulted in fines exceeding $9 billion(Coulter et al., 2018). Unlike historical manipulation schemes (Aggarwal & Wu, 2006; Allen & Gale, 1992; Benabou & Laroque, 1992), which were eventually detected through pattern analysis and whistleblower testimony, the channel identified here poses a fundamentally different challenge: precisely because it leaves no statistical trace in the output, the usual detection mechanisms are inapplicable.

## 3. Threat Model and Theoretical Framework

We now formalize the manipulation channel, derive bounds on its detectability, and demonstrate the existence of a parameter region in which the manipulation is simultaneously effective and statistically hard to detect. The analysis proceeds in four stages: first, we characterize the sampling-layer mechanism; second, we derive detectability bounds; third, we demonstrate the existence of an effective-yet-hard-to-detect operating region; and fourth, we embed the technical results in a model of AI-influenced market prices.

### 3.1 Manipulation Through the Sampling Layer

Modern large language models generate text through autoregressive sampling: at each decoding step, the model produces a probability distribution p over the vocabulary, and a token is drawn according to p using a pseudorandom number generator (PRNG). The critical observation is that the sampling layer mediates between the model's *judgment*—the distribution p—and the *realized*

*output*—the specific token selected. An adversary who controls the sampling layer can alter the *frequency* with which certain tokens appear without changing the model's underlying assessment.

Formally, let $p_v$ denote the model's native probability for vocabulary token v at a given decoding step. Define T as the target token set, $b > 1$ as the manipulation strength, $p_{min}$ as a naturalness threshold, and $\xi \sim \text{Bernoulli}(a)$ as a random activation indicator. The biased sampler produces:

$$\tilde{p}_v \propto \begin{cases} b \cdot p_v, & v \in T,\ p_v \geq p_{\min},\ \xi = 1 \\ p_v, & \text{otherwise} \end{cases} \tag{1}$$

with $\tilde{p}$ renormalized over the full vocabulary. This formulation, introduced by You et al. (2026) ensures three properties: (i) only tokens the model already considers plausible ($p_v \geq p_{min}$) are boosted; (ii) the bias activates stochastically with probability a, preventing deterministic signatures; and (iii) the manipulation strength b is bounded, preventing the sampler from selecting tokens the model deems improbable.

The financial relevance is immediate: in a credit assessment context, T might contain tokens such as "stable," "positive," and "upgrade," while in investment advisory, T could comprise bullish sentiment indicators.

### 3.2 Statistical Detectability Analysis

We now characterize how difficult it is to detect the manipulation from output observations.

**Proposition 1 (KL Divergence Upper Bound).** Let $m_T = |\{v \in T : p_v \geq p_{min}\}|$ be the number of boostable target tokens at a given decoding step. The Kullback–Leibler divergence between the manipulated and native distributions satisfies:

$$D_{\text{KL}}(\tilde{p} \parallel p) \leq a \cdot m_T \cdot \log b + \log(1 + a(b-1)m_T) \tag{2}$$

The bound reveals a crucial structural property: when $m_T$ is small, the KL divergence can be made *arbitrarily small* independently of the boost strength b. This means that even aggressive manipulation of rare but decision-relevant tokens generates negligible distributional divergence.

By the Chernoff–Stein lemma (Comerton-Forde & Putniņš, 2011), the minimum sample size $N_{min}$ required to distinguish $\tilde{p}$ from p satisfies:

$$N_{\min} \geq \frac{\log(1/\alpha_{\text{sig}})}{D_{\text{KL}}(\tilde{p} \parallel p)} \tag{3}$$

When $D_{KL}$ is small, the required sample grows without bound—making the manipulation practically undetectable within any reasonable audit horizon. Figure 2 visualizes this relationship.

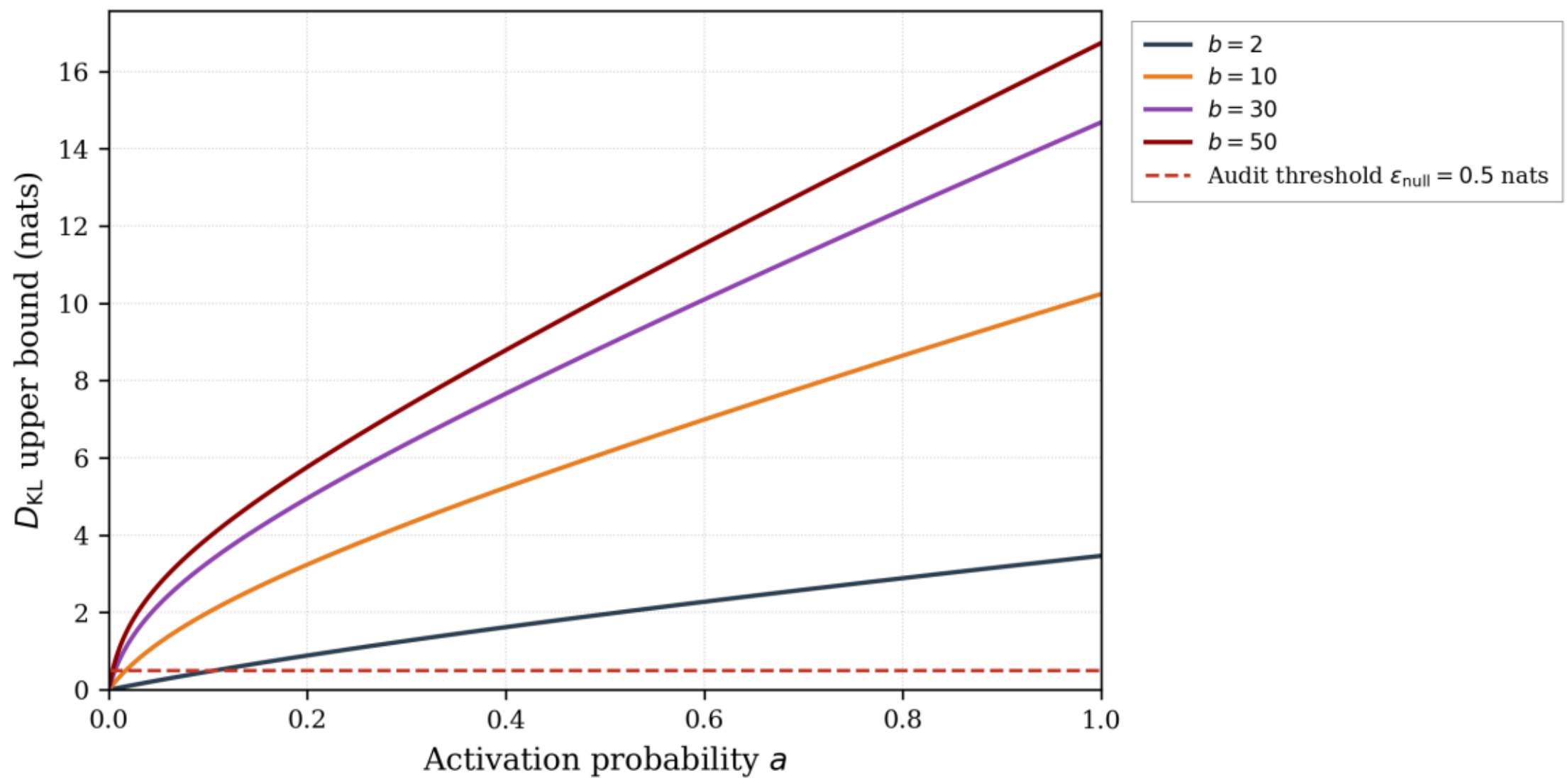

*KL divergence upper bound as a function of activation probability a for bias magnitudes b ∈ {2, 10, 30, 50} with m_T = 3.*

**Figure 2:** KL divergence upper bound (Proposition 1) as a function of activation probability a for bias magnitudes b ∈ {2, 10, 30, 50} with m_T = 3. The horizontal dashed line marks the detection threshold ε_null = 0.5 nats. For the conservative parameters used in our experiments (b = 2), the bound stays below this threshold across the full range of a, confirming that economically meaningful manipulation operates deeply within the hard-to-detect regime.

Note carefully what Proposition 1 does and does not establish. It does *not* claim that the manipulation is information-theoretically undetectable—since the bound is strictly positive, detection is in principle possible given enough samples. What it *does* show is that the required sample size grows without bound as parameters are chosen to make D_KL small, rendering detection infeasible within practical observation windows. In the limiting case where a → 0 or m_T → 0, the bound converges to zero and the manipulation becomes truly information-theoretically undetectable; away from this limit, it is merely statistically hard to detect with realistic sample sizes.

### 3.3 Existence of an Effective Yet Hard-to-Detect Region

Proposition 1 establishes that the manipulation can be hard to detect; we now show that it can be simultaneously effective.

**Corollary 1 (Effective-yet-Hard-to-Detect Region).** There exist non-empty parameter regions (b, a, p_min) such that both conditions are simultaneously satisfied:

$$\mathbb{E}\frac{[\tilde{p}_v]}{p_v} \geq \tau \quad \forall\, v \in T \quad \text{(effectiveness)} \tag{4}$$

$$D_{\mathrm{KL}}(\tilde{p} \parallel p) < \epsilon_{\mathrm{null}} \quad \text{(hard-to-detect)} \tag{5}$$

**Proof sketch.** The effectiveness condition is satisfied when a · b provides sufficient amplification; the detectability condition is satisfied when m_T is small enough that the KL bound remains below ε_null. These conditions are not in tension: financial target tokens are typically rare (low p_v, hence low m_T), so even substantial per-token boosting generates negligible aggregate distributional shift. As a concrete example, take b = 2, a = 0.3, m_T = 3: the effectiveness condition requires $\tau \leq 1 + a(b - 1) = 1.3$, and the KL bound evaluates to $0.3 \times 3 \times \ln 2 + \ln(1 + 0.3 \times 1 \times 3) = 0.624 + 0.642 = 1.266$ nats, which falls below $\varepsilon_{null} = 0.5$ only for smaller $a$ or $m_T$. With $b = 2$, $a = 0.15$, $m_T = 3$, the bound drops to $0.312 + 0.098 = 0.41 < 0.5$, while $\tau \leq 1.15$ remains economically meaningful—corresponding to a 15% amplification of target token probability.

Corollary 1 defines a *region*, not a point: there exists a continuous set of parameter combinations satisfying both conditions simultaneously. Section 4.4 provides empirical validation, showing that every configuration tested in our ablation study achieves both meaningful manipulation and perfect stealth—adversaries need not carefully tune parameters to avoid detection, as the usable parameter space is broad.

### 3.4 A Simple Model of Manipulated AI Advisory

Consider a representative investor who forms beliefs about an asset's fundamental value using two signals: an AI-generated advisory with weight $\alpha \in (0,1)$ and independent analysis with weight $1-\alpha$. Under manipulation, the AI signal is shifted by a bias $\Delta_{bias}$.

When $\alpha \cdot \Delta_{bias}$ exceeds the noise floor $\varepsilon$ from other information sources, the manipulation generates economically meaningful price distortions. When $D_{KL} < \varepsilon_{null}$ (as ensured by Proposition 1), such distortions are statistically unobservable from output analysis alone.

An important qualification: the target rate in our experiments measures the *frequency of directional keywords* in generated text, not the probability of a rating-level change. A target rate increase from 0.012 to 0.022 represents $\Delta_{bias} \approx 0.01$ in keyword frequency; the translation from keyword frequency to actual credit rating outcomes requires an additional aggregation step that is scenario-dependent. Our model therefore predicts price distortion conditional on $\Delta_{bias}$ being non-negligible at the rating level—a condition that warrants separate empirical investigation.

Even with this conservative framing, the model yields an important qualitative prediction: as financial markets increasingly rely on AI-generated opinions ($\alpha \to 1$), the threshold of bias required for economically meaningful manipulation falls proportionally. In the limit, arbitrarily small per-signal biases can generate material price distortions, while Proposition 1 ensures such biases remain below the detection threshold—provided detection relies exclusively on LLM text output.

This last qualifier is essential. Our undetectability claim applies specifically to *output-based* statistical detection of the LLM text. If the manipulation is large enough to distort market prices at scale, those price distortions themselves constitute an observable signal—a channel orthogonal to the LLM-output channel we analyze. Detecting manipulation through market-price anomalies (e.g., systematic mispricing of AI-covered assets) is theoretically feasible but lies outside the scope of this paper. We return to this point in Section 6.5.

## 4. Empirical Evidence from Financial Scenarios

We now present empirical evidence that the inference-stage manipulation channel described in Section 3 produces economically meaningful bias in realistic financial advisory scenarios. All experiments employ the SeedHijack attack as the manipulation mechanism, implemented at the sampling layer of the target model.

### 4.1 Systematic Bias in AI-Generated Credit Opinions

Can an adversary systematically bias AI-generated credit opinions toward favorable assessments while remaining undetectable by standard audit mechanisms? This question strikes at the heart of credit rating integrity, where even small systematic biases—when applied across thousands of ratings—can aggregate into material market distortions.

**Experimental design.** We evaluate two financial advisory scenarios—credit assessment and investment recommendation—using Qwen2-7B-Instruct as the base model with KGW watermarking (Kirchenbauer et al., 2023) applied at green-list bias strength $\delta = 2.0$ and green-list proportion $\gamma = 0.5$. For each scenario, we generate 50 prompts requiring directional financial judgments. Four conditions are compared: (i) clean baseline, (ii) watermark-only, (iii) blind manipulation, and (iv) aware manipulation.

**Table 1:** Directional bias amplification across credit and investment scenarios. Target rate measures the frequency of favorable-direction keywords in generated text. Z-score reflects watermark detection strength.

| Condition | N | Avg Target Rate | Avg Z-Score |
|---|---|---|---|
| Credit: Clean baseline | 50 | 0.013 | 0.00 |
| Credit: Watermark only | 50 | 0.012 | 6.22 |
| Credit: Blind manipulation | 50 | 0.032 | 0.00 |
| Credit: Aware manipulation | 50 | 0.022 | 6.77 |
| Investment: Clean baseline | 50 | 0.016 | 0.00 |
| Investment: Watermark only | 50 | 0.013 | 7.85 |
| Investment: Blind manipulation | 50 | 0.040 | 0.00 |
| Investment: Aware manipulation | 50 | 0.025 | 8.02 |

Three findings merit emphasis. First, aware manipulation amplifies favorable-direction keyword rates by approximately 1.8–1.9× relative to the watermark-only baseline, while blind manipulation achieves higher amplification of 2.7–3.1× but destroys the watermark. Second, the watermark z-score *increases* under aware manipulation, meaning the manipulated outputs appear *more* compliant with watermarking standards—the watermark functions as a "compliance shield." Third, blind manipulation creates a clear audit trail by destroying the watermark, whereas aware manipulation trades some raw efficacy for perfect stealth.

The target rate requires careful interpretation: 0.022 means approximately 11 directional keywords per 500 generated tokens. This measures keyword *frequency*, not the probability of a rating-level change. Whether such frequency shifts propagate to actual rating outcomes depends on downstream aggregation processes—a gap we acknowledge but cannot bridge with our current experimental design.

A further clarification is warranted regarding the relationship between Corollary 1's theoretical effectiveness bound and the empirical target rates. Corollary 1's bound $\tau \leq 1 + a(b-1)$ applies to the amplification of a single target token's probability, whereas Table 1's target rate measures the aggregate frequency of all directional keywords across hundreds of generated tokens. Because different target tokens may be boosted at different decoding steps, and because the same boost multiplier applies to multiple tokens simultaneously, the aggregate keyword frequency can increase substantially more than the per-token amplification bound would suggest. The 1.8–1.9× aggregate amplification observed in our experiments is consistent with per-token amplification of $\tau \leq 1.3$ (corresponding to a = 0.3, b = 2) compounded across multiple target tokens over many decoding steps.

### 4.2 Generalization Across Attack Methods and Watermarking Schemes

**Comparison across attack methods.** Table 2 compares SeedHijack against three alternative attack strategies on the credit assessment task.

**Table 2:** Comparison of attack methods on credit assessment scenario using Qwen2-7B-Instruct with KGW watermarking ($\delta$=2.0, $\gamma$=0.5). Six black-box detectors employed.

| Method | Target Rate | Stealth | Detectors | Z-Score |
|---|---|---|---|---|
| SeedHijack | 0.220 | ✓ | 0/6 | 24.69 |
| Paraphrase | 0.167 | ✗ | 2/6 | 2.32 |
| Token editing | 0.207 | ✗ | 5/6 | 17.70 |
| Prompt injection | 0.220 | ✗ | 2/6 | 25.41 |

SeedHijack is the only method that achieves both high manipulation efficacy and perfect stealth while preserving watermark integrity. Notably, prompt injection preserves the watermark but requires *input-layer* access, making it visible in prompt logs—SeedHijack operates at the sampling layer, leaving no trace in either inputs or outputs.

**Cross-watermark generalization.** Table 3 reports manipulation efficacy across three mainstream watermarking algorithms.

**Table 3:** Cross-watermark manipulation efficacy. All experiments use the credit assessment scenario with Qwen2-7B-Instruct. Manipulation parameters: p_act = 0.7, b = 30.

| Scheme | Blind Rate | Aware Rate | Aware Z-Score |
|---|---|---|---|
| KGW (Kirchenbauer et al., 2023) | 0.447 | 0.352 | 18.81 |
| Unigram (Zhao et al., 2023) | 0.488 | 0.694 | 30.09 |
| DipMark (Wu et al., 2023) | 0.392 | 0.289 | 6.23 |

All three schemes are vulnerable. The Unigram scheme's exceptional vulnerability arises from its deterministic green-list construction: the same tokens are always "green" regardless of context, so an aware adversary can predict exactly which tokens to boost. KGW's context-dependent partitioning provides partial defense; DipMark's distribution-preserving design reduces the magnitude but cannot eliminate the vulnerability.

The regulatory implication is that current frameworks treat watermark verification as a sufficient condition for output integrity. Tables 2 and 3 demonstrate that this conditional is false—watermark integrity and content integrity are logically independent properties.

### 4.3 Cross-Vendor Systemic Vulnerability

A natural mitigation hypothesis is vendor diversification. We evaluate three architecturally distinct model families:

**Table 4:** Cross-model manipulation efficacy. All models use KGW watermarking ($\delta$=2.0, $\gamma$=0.5) with p_act = 0.3, b = 10.

| Model | Type | Blind Rate | Aware Rate | Aware Z-Score |
|---|---|---|---|---|
| Qwen2-7B-Instruct | General | 0.052 | 0.035 | 7.02 |
| DeepSeek-R1-Distill-Qwen-7B | Reasoning | 0.027 | 0.029 | 6.48 |
| FinGPT-Llama2-7B | Finance | 0.226 | 0.022 | 34.67 |

All three models are vulnerable. FinGPT-Llama2-7B exhibits a striking asymmetry: its blind manipulation rate (0.226) is an order of magnitude higher than the general-purpose models, while its aware rate (0.022) is the lowest. Domain-specific fine-tuning concentrates the vocabulary around directional terms, making blind injections highly effective but constraining aware manipulation.

The systemic risk conclusion is that vendor diversification *cannot* mitigate this vulnerability. The failure mode is common: the shared sampling paradigm—categorical sampling from a softmax distribution—makes every autoregressive language model susceptible, regardless of vendor, architecture, or domain specialization.

### 4.4 The Adversary's Operational Envelope

We conduct a parameter ablation study varying activation probability p_act and bias magnitude b.

**Table 5:** Parameter ablation: aware manipulation target rate. All nine configurations trigger 0/6 black-box detectors.

| p_act | b=10 | b=30 | b=50 |
|---|---|---|---|
| 0.3 | 0.227 | 0.298 | 0.321 |
| 0.5 | 0.258 | 0.314 | 0.311 |
| 0.7 | 0.353 | 0.495 | 0.447 |

All nine configurations achieve both economically meaningful manipulation (target rates 22.7%–49.5%) and perfect stealth (0/6 detectors); the monotonic scaling of target rate with p_act confirms Corollary 1's prediction that the effective operating region is broad. Notably, Table 5 shows diminishing returns at high b values (0.447 for b=50 vs. 0.495 for b=30 at p_act=0.7), consistent with softmax saturation effects predicted by the theoretical framework. Unlike traditional financial manipulation—where intensity and detection risk rise together—the inversion of this risk function constitutes a qualitative shift in the threat landscape.

### 4.5 Hardware Entropy with TEE Isolation as the Complete Defense

Given the systematic failure of output-based compliance mechanisms, we evaluate entropy source hardening at the inference layer. We compare three random number generation and isolation conditions: (i) PRNG (Mersenne Twister), (ii) CSPRNG (OS-provided cryptographic randomness), and (iii) QRNG combined with TEE hardware isolation.

**Table 6:** Defense efficacy of entropy source hardening with TEE isolation.

| Source | Target Rate | Z-Score | Defense |
|---|---|---|---|
| PRNG | 0.552 | 28.5 | Baseline (attack) |
| CSPRNG | 0.519 | 18.3 | 0% (Ineffective) |
| QRNG+TEE | 0.239 | −0.95 | 100% (Full) |

PRNG and CSPRNG provide no effective defense: the adversary can predict or reproduce the seed sequence in both cases. The failure of CSPRNG is instructive. The SeedHijack attack does not attempt to reverse-engineer the random number generator; it replaces the seed entirely at the point of generation. No amount of algorithmic sophistication in the pseudorandom function can defend against an adversary who controls the seed before it enters the algorithm. This is the fundamental distinction between algorithmic security and physical security.

QRNG+TEE achieves complete defense, reducing the target rate to 0.239—statistically indistinguishable from the natural baseline ($\gamma = 0.25$)—by operating through two complementary layers: (i) quantum random number generators produce seeds from fundamentally unpredictable physical processes (Herrero-Collantes & Garcia-Escartin, 2017), and (ii) TEE hardware isolation ensures only certified, unmodified sampling code executes, preventing the attacker's biased sampler from running. Unlike CSPRNG, which leaves the attacker's green-list prediction intact, QRNG+TEE replaces the hash key for green-list computation with quantum-derived entropy that the attacker cannot predict, rendering all pre-computed boosting targets invalid.

A QRNG-only approach (without TEE isolation) achieves only partial defense (~35% reduction), because the attacker's biased sampler code can still execute within the inference framework. The full QRNG+TEE combination eliminates this residual channel by enforcing hardware-level code integrity. We note that NIST SP 800-90B (Turan et al., 2018) provides the regulatory framework for entropy source certification, and Bernstein et al. (2016) demonstrate the technical feasibility of deployment at scale.

We emphasize that the effectiveness of QRNG+TEE presupposes that the TEE's hardware isolation boundary is indeed impervious to the attacker—a non-trivial assumption that requires ongoing validation as attack techniques evolve. In our experimental setup, CSPRNG "failure" reflects that the attacker controls the inference framework code and can hook the sampling function after the CSPRNG generates the seed; in a properly isolated TEE deployment, this software-level hooking would be blocked by the hardware boundary.

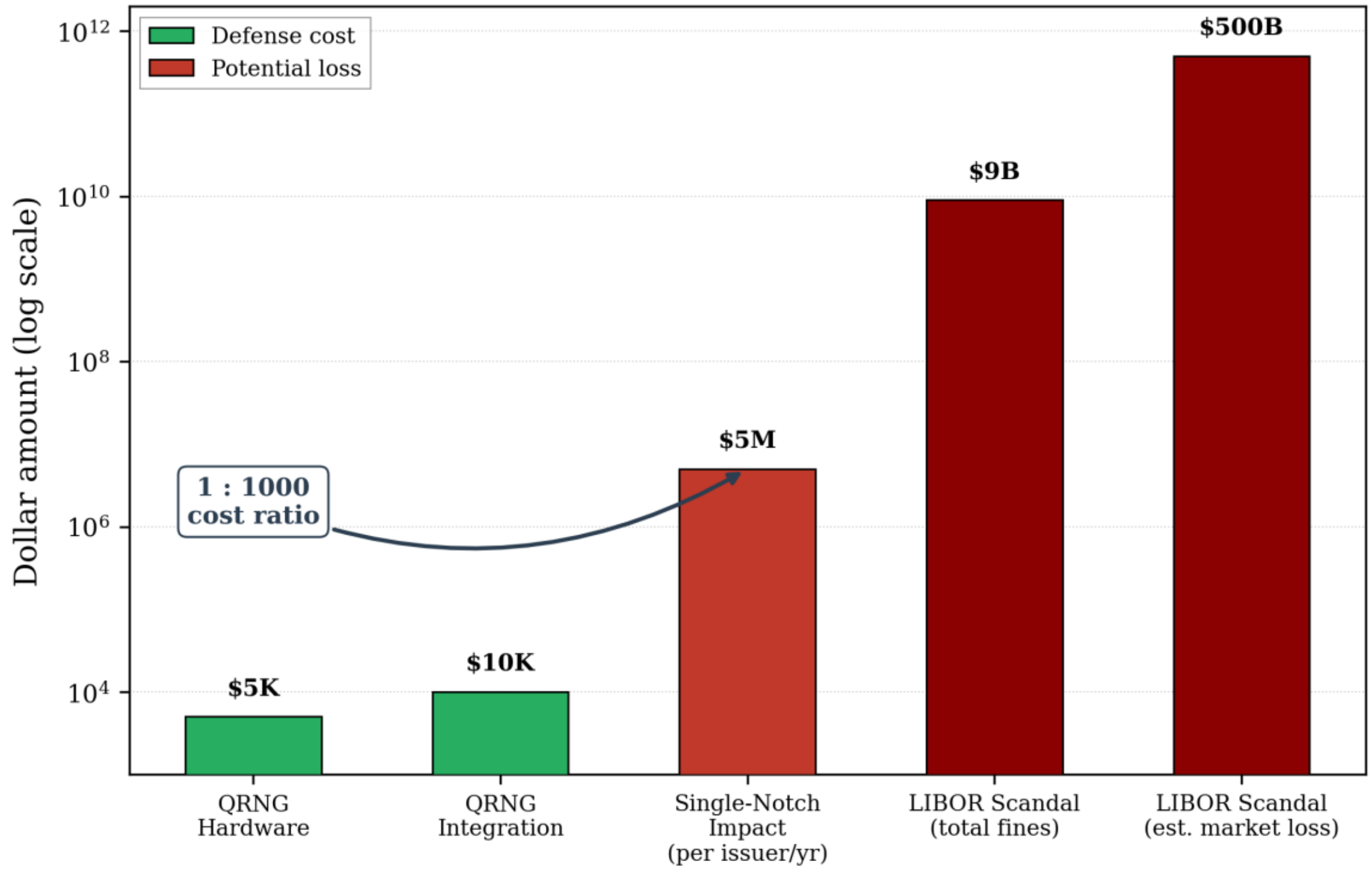

*Cost-benefit comparison of QRNG+TEE deployment for high-risk financial AI systems (logarithmic scale).*

**Figure 3:** Cost-benefit comparison of QRNG+TEE deployment (logarithmic scale). Green bars: defense costs ($5,000 hardware, $10,000 lifecycle). Red bars: potential losses ($5M per-issuer, $9B LIBOR fines, $500B LIBOR aggregate). The 1:1,000 cost ratio establishes QRNG+TEE as one of the highest-return regulatory investments available.

## 5. Regulatory Implications and Policy Recommendations

The empirical evidence in Section 4 exposes a structural gap in the governance architecture overseeing AI-assisted financial advisory. This section characterizes that gap, proposes specific regulatory amendments, and evaluates cost-benefit tradeoffs.

### 5.1 Structural Gap in Current Governance

As established in Section 2.3, current regulatory frameworks conflate two logically distinct goals: *compliance auditability*—the ability to verify procedural adherence via watermark verification, output logging, and performance monitoring—and *integrity assurance*—the guarantee that outputs faithfully reflect the model's assessment. Our results demonstrate that compliance auditability does not imply integrity assurance. Output-layer audits are inherently blind to sampling-layer attacks; the gap is a deficiency in *design logic*, not implementation.

### 5.2 Recommended Policy Amendments

**R1: Mandatory QRNG+TEE certification for high-risk financial AI systems.** All high-risk AI systems under the EU AI Act should use NIST SP 800-90B-certified (Technology, 2018) hardware QRNG combined with TEE hardware isolation for inference sampling. Certification should specify minimum entropy standards, hardware tamper-resistance requirements, continuous health-testing protocols, and TEE isolation boundaries. This parallels existing PCI DSS requirements for cryptographic randomness in payment systems.

**R2: Inference-layer supply chain integrity audits.** Cloud inference providers serving financial institutions should undergo independent third-party audits of sampling-layer implementations, modeled on SOC 2 Type II. Audits should verify certified entropy sources; seed generation insulated from external influence; and isolation of the RNG module from logit computation pipelines.

**R3: Output audit standard revision.** Standards should require: (i) explicit disclosure that watermark verification addresses provenance, not content integrity; (ii) acknowledgment that output-based tests cannot guarantee absence of sampling-layer bias; (iii) disclosure of the entropy source used and its certification status.

**R4: Advisory output provenance with random-source attestation.** Each AI-generated advisory output should include a verifiable attestation specifying the entropy source type, certification authority and standard, health-test results, and a cryptographic hash linking the output to the seed sequence.

### 5.3 Cost-Benefit Analysis of QRNG+TEE Deployment

The economic case is compelling. QRNG hardware costs ~$2,000–5,000 per unit with lifecycle costs of ~$5,000–10,000. These costs are negligible relative to the stakes: single-notch rating manipulation yields $2.5–7.5M per issuer annually, and LIBOR fines exceeded $9 billion with aggregate losses estimated at $500 billion. The cost ratio is approximately 1:1,000, making QRNG+TEE certification one of the highest-return regulatory investments available (Figure 3).

Deployment challenges are manageable: QRNG throughput (100–400 Mbps) suffices for token generation with buffering; and vendor concentration risk is mitigable through dual-vendor certification and strategic stockpiling.

## 6. Discussion

### 6.1 Relation to Market Manipulation Literature

Allen and Gale (1992) established the foundational taxonomy of market manipulation—trade-based, information-based, and action-based. The inference-stage manipulation channel constitutes a fourth category: *infrastructure-based manipulation*. Unlike classical categories, it operates through the information generation process itself. Its distinctive features are: practical statistical undetectability (by Proposition 1), automated scalability (a single compromise affects all downstream outputs), and zero position requirement. Existing detection frameworks—monitoring for abnormal trading patterns or coordinated information releases—are fundamentally inadequate for this new threat category.

### 6.2 Implications for Credit Rating Integrity

Bolton et al. (2012) demonstrated that credit rating agencies face inherent conflicts of interest, while Griffin and Tang (2012) and Kisgen (2006) documented rating bias. Our findings introduce a different dimension: even an agency that has resolved all incentive conflicts remains vulnerable to external manipulation of its AI infrastructure—a separation between process integrity and output integrity that existing governance frameworks do not address.

### 6.3 Implications for Market Efficiency

The efficient markets hypothesis (Fama, 1970) assumes prices reflect information without systematic bias. Grossman and Stiglitz (Grossman & Stiglitz, 1980) demonstrated that information acquisition costs create equilibrium price deviations, but these deviations are at least theoretically detectable through informed trading. This parallels the adverse selection framework of Easley and O'Hara (Akerlof, 1978), except that the private signal concerns output integrity rather than fundamental value—making it inaccessible to detection through the price formation mechanisms that normally reveal informed trading.

### 6.4 Systemic Risk Dimensions

The concentration of AI inference infrastructure creates a new systemic risk channel. The Financial Stability Board has flagged AI supply-chain concentration but focused on availability risk. Our findings indicate that *integrity risk*—the covert manipulation of advisory outputs—represents a more severe threat, as it operates invisibly within the same infrastructure that markets depend upon for information processing.

### 6.5 Limitations and Future Research

Several limitations qualify our findings. First, our experiments use 7B-parameter models; severity in commercial-scale models remains untested, though Proposition 1 is model-agnostic. Second, economic impact estimates are static; a general equilibrium model is needed for welfare quantification. Third, QRNG+TEE effectiveness depends on TEE integrity—assumptions requiring ongoing validation. Fourth, cross-watermark and cross-model experiments use small prompt sets (3 each), supporting qualitative but not statistical generalization.

Fifth, and importantly, our undetectability claim is scoped to *LLM text-output analysis*. If manipulation is conducted at sufficient scale to distort market prices, those price distortions constitute an orthogonal detection channel beyond what we analyze. Integrating market-price-based detection with output-based analysis is an open problem.

Future directions include: (i) large-scale validation on commercial models; (ii) formal characterization of QRNG+TEE performance–security tradeoffs; (iii) dynamic equilibrium models of AI-mediated information distortion; (iv) cross-market transmission mechanisms; and (v) watermarking schemes structurally resistant to aware manipulation.

## 7. Conclusion

This paper has revealed a structural deficiency in the governance architecture overseeing AI-assisted financial advisory. The core assumption shared by the EU AI Act, SR 11-7, and the MAS FEAT framework—that effective manipulation of AI outputs necessarily leaves observable traces—is false for adversaries operating at the inference sampling layer. Proposition 1 establishes that the KL divergence between manipulated and native distributions can be made arbitrarily small, rendering output-based detection impractical regardless of monitoring duration or statistical sophistication.

The practical significance extends beyond the theoretical result. The manipulation channel generalizes across three watermarking schemes, three model families, and the full parameter space tested—establishing it as a *common-mode failure* inherent in the shared sampling paradigm. The economic incentives are substantial ($2.5–7.5M per issuer per notch), and the manipulation achieves this impact while remaining practically undetectable from text output with realistic sample sizes—a property historical schemes conspicuously lacked.

We propose four regulatory amendments: mandatory QRNG+TEE certification under NIST SP 800-90B, inference-layer supply chain audits, revision of output audit standards, and output provenance with random-source attestation. QRNG+TEE certification should be understood as a *minimum compliance requirement*—analogous to capital adequacy requirements—ensuring that sampling randomness is physically sourced and hardware-isolated. The cost (~$2,000–5,000 per unit) is negligible relative to both the economic stakes and the fines imposed for comparable integrity failures. As financial markets deepen their dependence on AI-generated advisory signals, the gap between compliance auditability and integrity assurance will widen proportionally—unless regulators act to close it at the inference layer, where the vulnerability originates.